\documentclass[aps, reprint, nofootinbib, longbibliography, superscriptaddress, floatfix]{revtex4-1}

\usepackage{amsmath, amssymb, amsthm}
\usepackage{bm,bbm}

\usepackage{hyperref}
\usepackage[utf8]{inputenc}

\usepackage{soul}
\usepackage{url}
\usepackage{mathtools}
\usepackage{braket}
\usepackage{dsfont}
\usepackage[english]{babel}
\usepackage{graphicx} 
\usepackage{tikz}
\usetikzlibrary{quantikz}

\usepackage{hyperref}
\usepackage{natbib} 
\usepackage{physics}

\usepackage{listings}
\usepackage{xcolor}
\usepackage{amssymb}

\setbox0=\vbox to \vsize{\unvbox255\vfill}

\usepackage{multirow}
\usepackage{tabu}
\usepackage{makecell}
\usepackage{color}

\usepackage{lipsum,parskip,booktabs,graphicx,tabulary,tabularx}

\usepackage{wrapfig}
\newtheorem{definition}{Definition}

\newcommand{\LDIGUS}{
Lighthouse Disruptive Innovation Group, LLC,
7 Broadway Terrace, Apt 1,
Cambridge MA 02139,
Middlesex County, Massachusetts (USA)
}

\newcommand{\LDIGEU}{
Lighthouse Disruptive Innovation Group Europe, SL.
Barcelona - Spain
}

\newcommand{\DSD}{
Smart Society Research Group -
La Salle - Universitat Ramon Llull,
Carrer de Sant Joan de La Salle, 42,
08022 Barcelona (Spain)
}

\newcommand{\MIT}{
MIT Media Lab - City Science Group, Cambridge, USA
}

\usepackage{babel}

\graphicspath{{./images/}}

\begin{document}
\title{Efficient Quantum Modular Arithmetics for the ISQ Era}

\author{Parfait Atchade-Adelomou}
\affiliation{\MIT}
\email{parfait@mit.edu}
\affiliation{\DSD}
\affiliation{\LDIGUS}

\email{parfait.atchade@lighthouse-dig.com}


\author{Saul Gonzalez}
\affiliation{\LDIGEU}
\email{saul.gonzalez@lighthouse-dig.com}

\date{November 2023}

\begin{abstract}
As we venture into the Intermediate-Scale Quantum (ISQ) era, the proficiency of modular arithmetic operations becomes pivotal for advancing quantum cryptographic algorithms. This study presents an array of quantum circuits, each precision-engineered for modular arithmetic functions critical to cryptographic applications. Central to our exposition are quantum modular adders, multipliers, and exponential operators, whose designs are rigorously optimized for ISQ devices. We provide a theoretical framework and practical implementations in the PennyLane quantum software, bridging the gap between conceptual and applied quantum computing. Our simulations validate the efficacy of these methodologies, offering a strategic compass for developing quantum algorithms that align with the rapid progression of quantum technology.


\end{abstract}
\maketitle

\section{Introduction}\label{sec:intro}
Quantum computing, with its transformative potential, is driving a paradigm shift in computational methodologies. Classical operations foundational to traditional algorithms undergo scrutiny and reinvention as they transition into the quantum landscape. Modular arithmetic operators, specifically addition and multiplication, are quintessential due to their ubiquitous presence across classical and quantum algorithms. However, translating these operations to quantum mechanisms introduces many challenges, primarily because quantum operations necessitate reversibility. 

A notable constraint in the quantum programming paradigm is the stipulation for operations to be bijective when implemented \textit{Inplace}. This poses tangible impediments: for instance, iterative use of \textit{Outplace} operators in an algorithm demands auxiliary qubit registers for each iteration. The accumulating demand for such registers can quickly surpass the qubit capabilities of contemporary quantum devices. Given the centrality of modular arithmetic in seminal quantum algorithms, including Shor's factorization \cite{shor1994algorithms}, there is a pressing need to craft efficient, bijective modular operations.

This work embarks on an exploration and optimization of modular arithmetic operators tailored for the impending \textit{Intermediate-Scale Quantum (ISQ)} era \cite{FromNISQtoISQs}. Our endeavor is twofold: to dissect existing quantum modular operators, unearthing their intricacies, and to propose their fresh implementation, streamlined methodologies apt for the \textit{ISQ} landscape. By doing so, we aspire to fortify the bridge between theoretical quantum constructs and their tangible applications, equipping neophytes and seasoned quantum researchers with state-of-the-art modular arithmetic tools.

The organization of this manuscript is crafted to facilitate a comprehensive understanding of the research presented. Section~\ref{sec:motivation} articulates the impetus and intrinsic motivations guiding this inquiry. Section~\ref{sec:Preliminaries} constructs the conceptual framework, defining the terminology and foundational principles critical to the quantum modular arithmetic operators discourse. Section~\ref{sec:relatedwork} contextualizes our work within the existing scholarly landscape, critically examining preceding studies on quantum modular operators. The methodology, with its rigorous and detailed formulation, is expounded in Section~\ref{sec:IMPLEMENT}. After the theoretical exposition, Section~\ref{sec:results} validates our theoretical constructs with empirical evidence and provides a repository of executable code optimized for quantum computational platforms. The manuscript reaches its denouement in Section~\ref{sec:conclusions}, where we encapsulate our seminal contributions and delineate prospective trajectories for future scholarly exploration.

\section{Motivation and Problem Statement}\label{sec:motivation}
Quantum computing, while heralding unparalleled computational prowess, is still in its embryonic phase and is riddled with challenges. Among these is the constrained availability of qubits in nascent quantum architectures \cite{PhDParfait}. This context provides the foundation for our exploration.

The academic and research landscapes are replete with groundbreaking work \cite{shor1999polynomial,shor1994algorithms,Adr20,cerezo2021variational,ParfaitVQE,ParfaitAtchade, gonzalez2022gps, atchadeadelomou2023fourier} in quantum algorithms and their applications \cite{a14070194,consulpacareu2023quantum,atchade2022quantum, alonsolinaje2021eva,adelomou2020formulation,adelomou2023quantum,adelomou2020using}. However, despite this abundance, there is a conspicuous gap: a comprehensive synthesis of optimal algorithms, particularly in modular arithmetic. Furthermore, even amongst those that touch upon the subject, there needs to be more literature that delineates the most efficient implementations of modular arithmetic operators \cite{shor1999polynomial,rines2018high}. This observed lacuna, juxtaposed with the paramount significance of modular arithmetic in quantum algorithms, particularly in foundational ones like Shor's, accentuates the pressing need for a systematic investigation into this domain.

Driven by this palpable gap and a commitment to pragmatism and efficiency, we focus on devising modular arithmetic operators meticulously crafted for the \textit{ISQ} era. The quantum milieu of today necessitates algorithms that judiciously economize on qubit usage while embodying resilience and extensibility. By optimizing the utility of extant, bounded qubit resources, such algorithms should be congruent with contemporary quantum architectures and adaptable to the more sophisticated quantum machines on the horizon.

We aim to address this literature void by proposing optimal modular arithmetic operators and detailing their efficient implementations. While the implementations herein are tailored for PennyLane \cite{bergholm2022pennylane}, lauded for its flexibility and ease of integration, our methods' principles are designed with cross-platform compatibility. We intend to furnish the quantum research community with an authoritative resource on modular arithmetic that can be readily adapted and applied within diverse quantum computing frameworks, thereby fostering innovation in numerous quantum domains.

\section{Work Context}\label{sec:relatedwork}

Quantum computing, though nascent, has seen a surge of innovations tailored to harness the singular capabilities of quantum architectures. Central to this ecosystem is modular arithmetic, a linchpin in quantum cryptographic endeavors and quintessential to Shor's integer factorization algorithm \cite{shor1999polynomial}. 

As the field edges closer to the \textit{ISQ} era, the imperative to scrutinize and refine existing algorithms grows stronger. Modular exponentiation \cite{dimitrov1998algorithm,van2005fast,hong1996new} exemplifies this, melding classical and quantum paradigms \cite{gordon1998survey, marwaha2013comparative}. Noteworthy architectures like the Vedral-Barenco-Ekert (VBE) \cite{vedral1996quantum} and the Beckman-Chari-Devabhaktuni-Preskill (BCDP) \cite{beckman1996efficient} have set benchmarks. However, designs like Gossett's \cite{gossett1998quantum} and Beauregard's approach \cite{beauregard2003circuit}, championing pure quantum arithmetic \cite{ruiz2017quantum,GuilleQuantumArithmetics,Ket.G}, underscore the diversity and richness of the field.

In the rapidly evolving landscape of \textit{ISQ} computing, our objective is clear-cut: to craft modular arithmetic operators that resonate with current research and are poised for future advancements. While our solutions are intricately crafted upon the PennyLane framework, they are not bound by it. We have designed our algorithms to transcend platform-specific constraints, ensuring they can be effortlessly transposed to other quantum computing environments like Qiskit \cite{wille2019ibm}. The forthcoming sections will articulate the nuances of our modular arithmetic approach, juxtaposing it with existing paradigms and underscoring its pivotal role in the onward march of the  \textit{ISQ} era.

\section{Preliminaries}\label{sec:Preliminaries}

This section lays the foundational terminology and concepts pivotal to our study of quantum modular arithmetic operators. We establish the encoding scheme used throughout this work, which is critical for consistency in our theoretical and practical applications. Discussing the \textit{ISQ} era contextualizes our study within the current quantum computing landscape. Following this, we precisely classify operators into \textit{Inplace} and \textit{Outplace} categories, each with distinctive operational characteristics crucial for quantum algorithm design. We introduce our operators with formal definitions that serve as a precursor to the in-depth discussions in later sections.

\subsection{Notation and Qubit Encoding}\label{subsec:notation_encoding}

In alignment with the encoding standards of the PennyLane quantum computing framework \cite{bergholm2022pennylane}, we adopt the conventional big-endian format \cite{james1990multiplexed}.

\begin{definition}[Qubit Encoding in PennyLane]
PennyLane adheres to a big-endian qubit encoding convention, aligning with standard practices in quantum computation frameworks. Within this convention, \( a_0 \) is designated as the most significant bit (\textit{MSB}), consistently represented at the topmost or leftmost in quantum circuit diagrams, which corresponds to the highest order of magnitude in the binary representation. In contrast, \( a_{n-1} \) is the least significant bit (\textit{LSB}), placed at the bottommost or rightmost, representing the lowest order of magnitude. This encoding ensures coherence with PennyLane's infrastructure and promotes uniformity in implementing quantum algorithms. 
\end{definition}

For illustration, consider the integer \( a \) expressed in binary form as \( a = a_0 \cdot 2^n + a_1 \cdot 2^{n-1} + \ldots + a_n \cdot 2^0 \). When representing the quantum state corresponding to the decimal number \( 6 \) within PennyLane, the binary equivalent \( 110 \) is denoted as \(\ket{a_0a_1a_2} = \ket{110}\). This notation maps the binary representation such that \( a_0 \) is the \textit{MSB} and \( a_n \) is the \textit{LSB}, reflecting a left-to-right orientation in circuit schematics.

This encoding scheme is essential for adequately interpreting quantum registers and analyzing computational results, necessitating its comprehension for an accurate understanding of the quantum operations presented herein. Users employing alternative platforms, such as \textit{Qiskit}, should adjust the qubit ordering accordingly to maintain the integrity of the computational results.

\subsection{\textit{ISQ} era and Operator Classification}\label{subsec:operator_classification}

With the advent of the \textit{ISQ} era \cite{FromNISQtoISQs}, a refined comprehension of quantum operators gained paramount importance.

\begin{definition}[Intermediate-Scale Quantum (ISQ) Era]
The ISQ era bridges the gap between the Noisy Intermediate-Scale Quantum (NISQ) phase and the advent of fully fault-tolerant quantum computing. ISQ devices, benefiting from nascent quantum error correction techniques, enable extended quantum circuit depths, albeit within a limited qubit framework. This period emphasizes the fine-tuning of quantum algorithms to resonate with these newfound capabilities, heralding a new epoch of reliable quantum computations.
\end{definition}

Our study categorizes these operators into two fundamental types—\textit{Inplace} and \textit{Outplace}—which are instrumental for executing quantum arithmetic operations. These classifications facilitate a structured approach to algorithmic design and underpin the advancement of quantum computational methodologies.

\begin{definition}[Inplace Operator]
An operator is termed \textit{Inplace} if it applies a function \( f: \mathbb{N} \rightarrow \mathbb{N} \) directly to the input qubit register without the need for additional output space. For the associated quantum operator \( U_f \), the transformation is described as \( U_f\ket{x} = \ket{f(x)} \), where the output \( f(x) \) overwrites the original input \( x \) in the same register. The feasibility of an \textit{Inplace} operator hinges on the bijectivity of the function \( f \), which guarantees the reversibility of the operation, a requirement for quantum computations. 
\end{definition}
An illustrative example within this context is the modular addition by a constant \( k \) modulo \( N \), defined as \( Add_{\text{in}}(k, N) \), which computes \( \ket{x} \rightarrow \ket{x + k \mod N} \) using the same qubits that initially represented \( \ket{x} \).

\begin{definition}[Outplace Operator]
An \textit{Outplace} operator for a function \( f: \mathbb{N} \rightarrow \mathbb{N} \), and its corresponding quantum operator \( U_f \), operates on two distinct registers. It preserves the original input register \( |x\rangle \) intact and conveys the function's output to an auxiliary register set to \( |0\rangle \), resulting in the transformation \( U_f|x\rangle|0\rangle = |x\rangle|f(x)\rangle \). This design allows for the execution of non-bijective functions within a quantum framework. 
\end{definition}
For example, the operator \( Add_{\text{out}}(k, N) \) showcases this architecture by computing \( Add_{\text{out}}(k, N)\ket{a}\ket{0} \rightarrow \ket{a}\ket{a+k \mod N} \), where the auxiliary register initially in the state \( |0\rangle \) is employed to hold the output of the modular addition.

The operators analyzed in this study are:

\begin{itemize}
    \item \(Add_{\text{in}}(k,N)\): An \textit{Inplace} modular adder for the classical constant \(k\), operating modulo \(N\).
    \item \(Add_{\text{out}}(k,N)\): Its \textit{Outplace} counterpart for modular addition.
    \item \(Add_{\text{out}}(N)\): An \textit{Outplace} adder for dual quantum variables.
    \item \(Add_{\text{in}}(N)\): The \textit{Inplace} version for two quantum variables.
    \item \(Mult_{\text{out}}(k,N)\): \textit{Outplace} modular multiplier with the classical constant \(k\), modulo \(N\).
    \item \(Mult_{\text{in}}(k,N)\): \textit{Inplace} modular multiplier for the classical variable \(k\), modulo \(N\).
    \item \(Mult_{\text{out}}(N)\): \textit{Outplace} modular multiplier for two quantum variables.
    \item \(Exp_{\text{out}}(a,N)\): An \textit{Outplace} exponential operator.
\end{itemize}

By distinctly listing each operator and specifying its type, we ensure clarity and provide a detailed overview that serves as a foundation for the quantum computational principles discussed subsequently.

\section{Implementation of Quantum Modular Operators}\label{sec:IMPLEMENT}

The quantum modular operators, while foundational, enable a myriad of quantum algorithms, especially those focusing on cryptographic tasks and number-theoretic challenges.

\subsection{Fundamental Auxiliary Operators}\label{sec:auxiliary}

Our journey in crafting optimal modular arithmetic operators frequently intersects with the necessity for auxiliary quantum operators. These operators simplify our primary implementations and enhance their efficiency and adaptability, especially in the context of the \textit{ISQ} era. In this section, we spotlight these indispensable components:

\begin{itemize}
    \item \textbf{Quantum Fourier Transform (QFT)} \cite{coppersmith2002approximate}: The QFT is a linchpin for in-place qubit addition. Its transformation, as represented in equation \eqref{eq:QFT}, illuminates its relevance in the set of operators we delineate:
    \begin{align}
        QFT\ket{j_1}\ket{j_2}...\ket{j_n} \rightarrow \frac{1}{2^{n/2}} \Bigg[ 
        &(\ket{0}+\exp^{2\pi i 0.j_n}) \nonumber \\
        &\otimes (\ket{0}+\exp^{2\pi i 0.j_{n-1}j_{n}}) \nonumber \\
        &\otimes \ldots \nonumber \\
        &\otimes (\ket{0}+\exp^{2\pi i 0.j_{1}j_{2}...j_{n}}) \Bigg].
        \label{eq:QFT}
    \end{align}
    
   \item \textbf{Quantum Fourier Adder} \cite{GuilleQuantumArithmetics,ruiz2017quantum}: Central to quantum modular arithmetic is the \( Sum(k) \) operator, which facilitates the addition of a classical constant \(k\) to a quantum register state \(\ket{a}\). This operation leverages the transformative capabilities of the Quantum Fourier Transform (QFT) to transmute the addition into a sequence of controlled phase shifts. Initially, the quantum state undergoes the QFT, mapping the computational basis to a superposition where each component's phase is modulated by \(k\) (Eq. \eqref{eq:qft_adder}). Subsequently, the inverse QFT (QFT\(^{-1}\)) is applied, effectively synthesizing the augmented quantum state \(\ket{a + k}\). The operation is mathematically represented as:

\begin{equation}
\begin{split}
    \ket{a} \xrightarrow{\text{QFT}}
    & \frac{1}{2^{n/2}}\left(\sum_{p=0}^{2^n-1} e^{2\pi iap/2^n}\ket{p}\right) \\
    & \xrightarrow{Sum(k)}
    \frac{1}{2^{n/2}}\left( \sum_{p=0}^{2^n-1}e^{2\pi i(a+k)p/2^n}\ket{p}\right) \\
    & \xrightarrow{\text{QFT}^{-1}}
    \ket{a + k}.
\end{split}
\label{eq:qft_adder}
\end{equation}

The circuit implementation of this adder is illustrated in Fig. \eqref{fig:integer_add}, which integrates the QFT with controlled phase rotations to enable quantum addition, as elaborated in Eq. \eqref{eq:qft_adder}.

    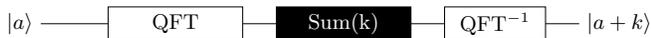
\begin{figure}[!h]
        \centering
        \resizebox{0.5\textwidth}{!}{
        \begin{tikzpicture}
            \draw (0,0) -- (8,0);
            
            \fill[white] (1,-0.25) rectangle (3,0.25);
            \draw (1,-0.25) rectangle (3,0.25) node[midway] {QFT};
            
            \fill[black] (3.5,-0.25) rectangle (5.5,0.25);
            \draw (3.5,-0.25) rectangle (5.5,0.25) node[midway, text=white] {Sum(k)};
            
            \fill[white] (6,-0.25) rectangle (7.5,0.25);
            \draw (6,-0.25) rectangle (7.5,0.25) node[midway] {QFT\(^{-1}\)};
            
            \node[anchor=east] at (0,0) {\(\ket{a}\)};
            \node[anchor=west] at (8,0) {\(\ket{a+k}\)};
        \end{tikzpicture}
        }
        \caption{Visualization of the integer quantum adder operator. }
        \label{fig:integer_add}
    \end{figure}
\end{itemize}

\begin{figure*}
    \includegraphics[width=1\textwidth]{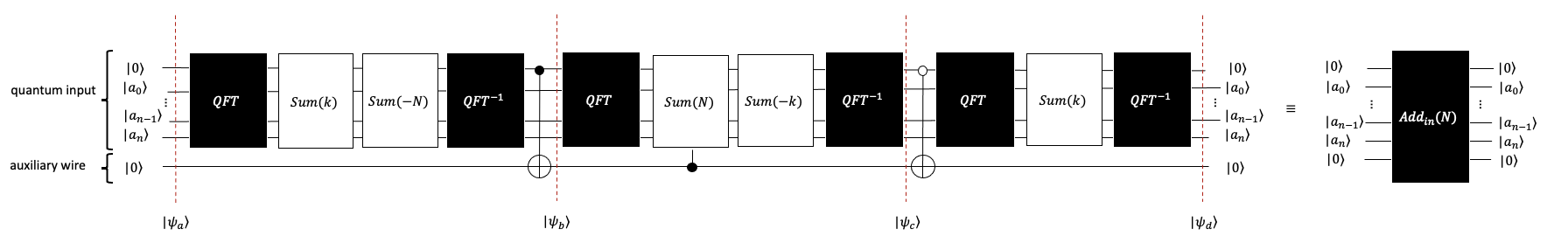}
\caption{Schematic of the \textit{Inplace} Modular Quantum-Classical Adder. Commencing with \(\psi_{a} = \ket{0}\ket{a}\ket{0}\), the circuit employs a dual auxiliary qubit system to facilitate underflow control, with one qubit corresponding to the MSB and another to the LSB. Upon addition of the negative modulus \(-N\) and subsequent comparison, if the condition \(a+k-N < 0\) is met, the terminal qubit is activated, yielding state \(\psi_{b}=\ket{a+k-N}\ket{b}\) for the definition of \( b \), refer to Eq.~\eqref{eq:def_b}). A conditional restoration of \(N\) follows for states satisfying \(a+k-N < 0\), effectuating the state \(\psi_{c} = \ket{a+(b-1)N}\ket{b}\). The final addition of \(k\) to the conditioned state \(\ket{a+(b-1)N}\ket{0}\) culminates in the modular sum \(\psi_{d}=\ket{a+k \mod N}\). Incorporating the \textit{QFT} within this process facilitates the efficient computation of addition and subtraction within the Fourier space, illustrating the sophisticated interplay between classical and quantum computational paradigms.}
\label{fig:sum_a_k_mod_N}
\end{figure*}

The ensuing segments delve deeper into these auxiliary operators' circuit designs and functionalities, elucidating their synergy with the primary modular arithmetic operations central to our discourse.

\subsection{\textit{Inplace} Modular Quantum-Classical Adder}\label{sec:inplace_adder}

Central to our exploration is the \textit{Inplace}  Modular Quantum-Classical Adder, represented by the equation \eqref{eq:for_U1} and the figure \eqref{fig:sum_a_k_mod_N}.
    \begin{equation}
        Add_{\text{in}}(k, N)\ket{a} \rightarrow \ket{a+k \mod N}.
        \label{eq:for_U1}
    \end{equation}

This operator facilitates the modular addition of a constant \(k\) to a qubit \(\ket{a}\) that resides in the Fourier basis. Given its foundational role, several other modular arithmetic operators are derived, including sums, multiplications, and even the exponential.

To construct this operator, we introduce the variable \(b\), determined as:
\begin{equation}
    b = 
    \begin{cases} 
        0 & \text{if } a+k \geq N, \\
        1 & \text{if } a+k < N.
    \end{cases}
    \label{eq:def_b}
\end{equation}

Here, \(b\) indicates if a modular adjustment is required:
\begin{itemize}
    \item \( b = 0 \): The sum surpasses the modulus, necessitating subtraction by \(N\).
    \item \( b = 1 \): The sum remains within modular constraints, and no adjustment is needed.
\end{itemize}

With \(b\) defined, our objective is to design a circuit effecting the transformation \(f(\ket{a}):=\ket{a+k+(b-1)N}\). This circuit (detailed in Fig. \eqref{fig:sum_a_k_mod_N}) comprises three primary steps:
\begin{enumerate}
    \item \textbf{Phase 1:} Assess if subtraction by \(N\) is required. If so, utilize an auxiliary qubit to transition to the state \(\ket{a+(b-1)N}\).
    \item \textbf{Phase 2:} Clean the auxiliary qubit.
    \item \textbf{Phase 3:} Add \(k\) to the resultant state from Phase 1, post-auxiliary qubit cleaning.
\end{enumerate}

For this work, it is essential to highlight that all operations and representations are considered within the standard computational basis.

\subsection{\textit{Outplace} Modular Quantum-Classical Adder}\label{sec:outplace_q_c_adder}

\begin{figure}
    \centering
    \resizebox{.5\textwidth}{!}{
    \includegraphics[width=1\textwidth]{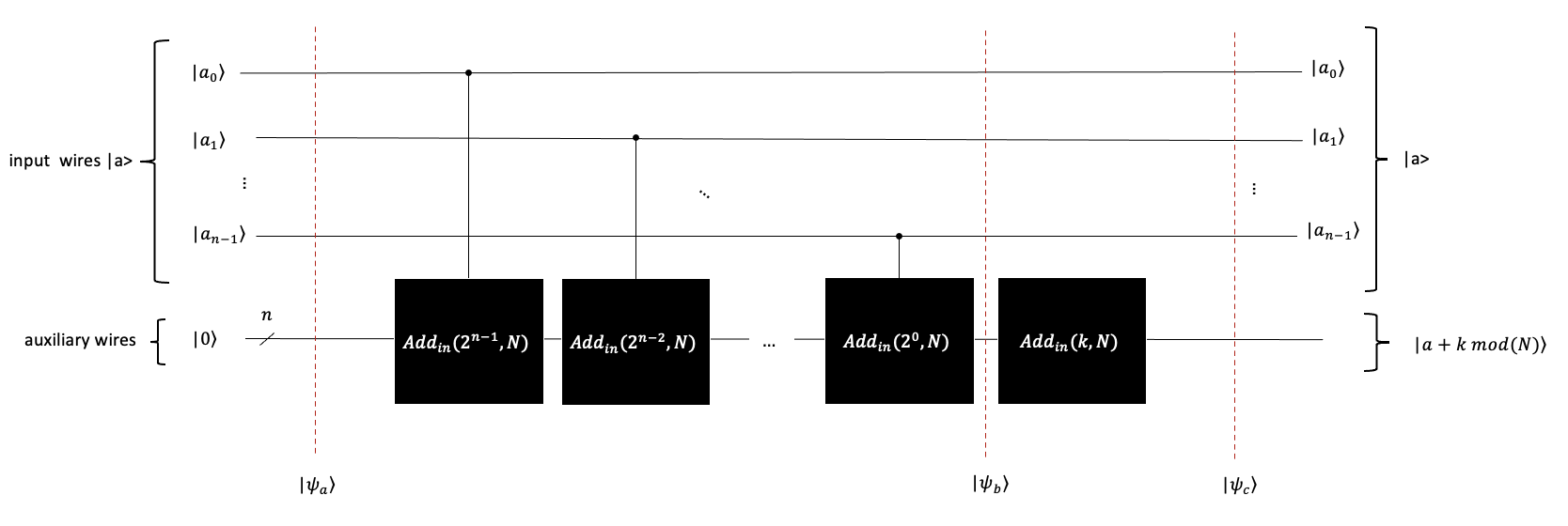}
    }
   \caption{Schematic of the \textit{Outplace} Modular Quantum-Classical Adder Operator. The operation initiates with the quantum register $\psi_{a} = \ket{a}\ket{0}$, where $\ket{a}$ encodes the quantum representation of the integer to which the classical constant $k$ will be added. Through the controlled application of the \(Add_{\text{in}}(N)\) operators, the system transitions to the intermediate state \(\psi_{b}=\ket{a}\ket{a}\). Subsequently, the circuit performs the modular addition, progressing to the state \(\psi_{c}=\ket{a}\ket{a + k \mod{N}}\), thereby encapsulating the sum in the second register. 
   }

    \label{fig:quantum_classical_outplace_sum}
\end{figure}

The \textit{Outplace} Modular Quantum-Classical Adder (see Fig. \eqref{fig:quantum_classical_outplace_sum}), denoted as \(Add_{\text{out}}(k,N)\), is defined by the transformation in equation \eqref{eq:add_out_q_c}.
\begin{equation}
Add_{\text{out}}(k,N)\ket{a}\ket{0} \rightarrow \ket{a}\ket{a+k \mod N}.
\label{eq:add_out_q_c}
\end{equation}

Leveraging the previously defined \(Add_{\text{in}}(k,N)\) operator, the operational sequence of \(Add_{\text{out}}(k,N)\) can be depicted as in equation \eqref{eq:add_out_q_c_steps}:
\begin{equation}
    \ket{a}\ket{0} \xrightarrow{\text{Copy}} \ket{a}\ket{a} \xrightarrow{Add_{\text{in}}(k,N)} \ket{a}\ket{a+k \mod N}.
    \label{eq:add_out_q_c_steps}
\end{equation}
Initially, the state \(\ket{a}\ket{0}\) is duplicated. Following this, the \(Add_{\text{in}}(k, N)\) operator is employed to yield the desired outcome.

\subsection{\textit{Inplace} Modular Quantum-Quantum Adder}\label{sec:inplace_q_q_adder}

\begin{figure}[!ht]
    \centering
    \resizebox{.5\textwidth}{!}{
    \includegraphics[width=1\textwidth]{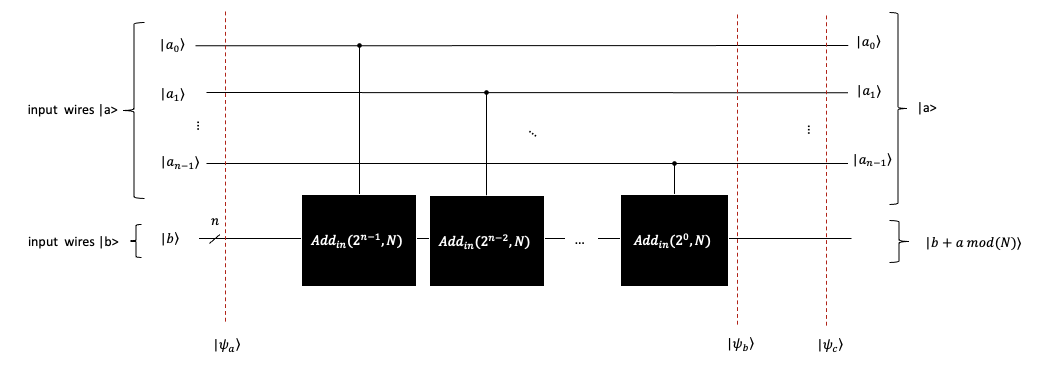}
    }
    \caption{Schematic of the \textit{Inplace} Modular Quantum-Quantum Adder Circuit. The process begins with the state \(\psi_{a} = \ket{a}\ket{b}\), where both \(\ket{a}\) and \(\ket{b}\) are quantum representations of the integers to be combined. The circuit executes the addition \textit{Inplace}, such that the second register transitions to \(\psi_{b} = \ket{a}\ket{a + b \mod{N}}\), encapsulating the sum within the existing qubit structure. The \textit{Inplace} operation preserves the initial quantum register \(\ket{a}\), while the second register is updated to reflect the modular addition.}
    \label{fig:quantum_quantum_inplace_sum}
\end{figure}

The \textit{Inplace} Modular Quantum-Quantum Adder (see Fig. \eqref{fig:quantum_quantum_inplace_sum}), symbolized as \(Add_{\text{in}}(N)\), performs the transformation outlined in equation \eqref{eq:adder_in_q_q}:
\begin{equation}
    Add_{\text{in}}(N)\ket{a}\ket{b} \rightarrow \ket{a}\ket{a+b \mod N}.
    \label{eq:adder_in_q_q}
\end{equation}

The operator's operational efficacy in isolation necessitates transposing input states into the Fourier basis via the Quantum Fourier Transform (QFT). However, when this operator is integrated with our \( \text{Add}_{\text{in}}(k,N) \) in a composite quantum circuit, the initial QFT and the concluding inverse QFT (\( QFT^{-1} \)) operations are unnecessaries. This optimization reduces the quantum resource requirements associated with Fourier basis transformations. Adopting this integrated approach, we capitalize on the standard configuration of the computational basis to implement controlled additions, where each qubit \( \ket{a_i} \) in the register \( \ket{a} \) contributes proportionally to its binary significance, as clarified in Equation \eqref{eq:ctrl_adder_in_q_q}.

\begin{equation}
    c_{\ket{a_i}}-Add_{\text{in}}(2^{n-i-1}, N) \ket{b},
    \label{eq:ctrl_adder_in_q_q}
\end{equation}
wherein \(c_{a_i}\) designates the control derived from qubit \(\ket{a_i}\). 

The associated quantum circuit is showcased in Fig. \eqref{fig:quantum_quantum_inplace_sum}.

\subsection{\textit{Outplace} Modular Quantum-Quantum Adder}\label{sec:outplace_q_q_adder}

The \textit{Outplace} Modular Quantum-Quantum Adder, referred to as \(Add_{\text{out}}(N)\), is characterized by the transformation described in equation \eqref{eq:outplace_q_q}.
\begin{equation}
    Add_{\text{out}}(N)\ket{a}\ket{b}\ket{0} \rightarrow \ket{a}\ket{b}\ket{a+b \mod N}.
    \label{eq:outplace_q_q}
\end{equation}

\begin{figure}
    \centering
    \resizebox{.5\textwidth}{!}{
    \includegraphics[width=1\textwidth]{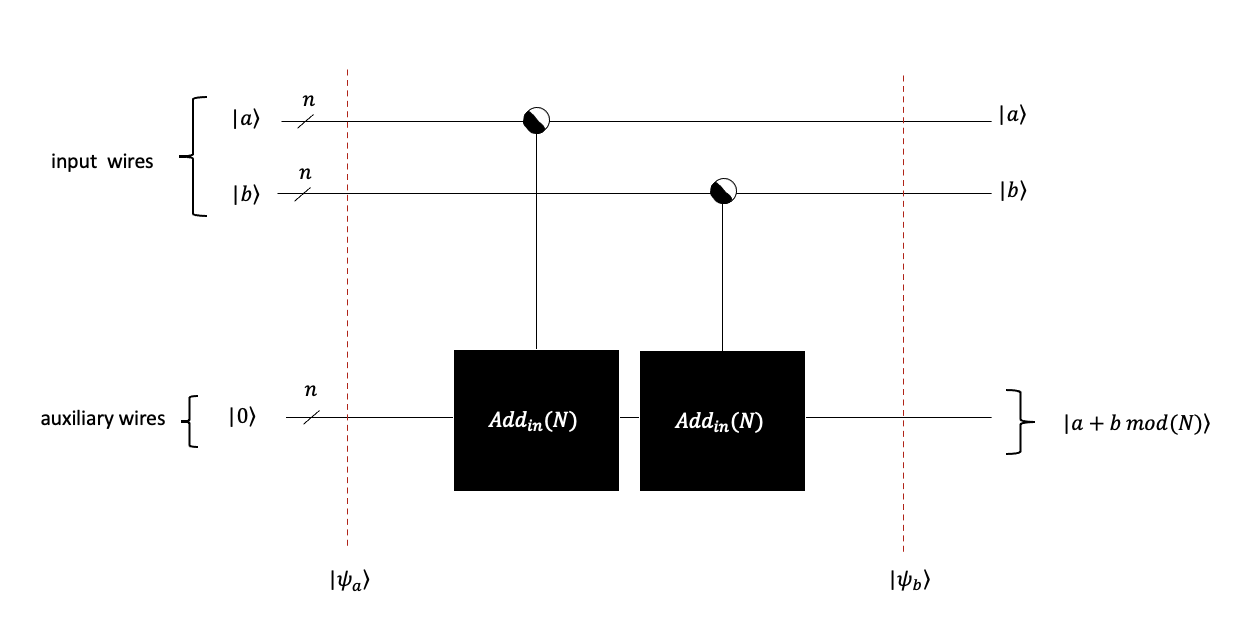}
    }
    \caption{Depiction of the \textit{Outplace} Modular Quantum-Quantum Adder Operator. The circuit commences with two quantum registers in the state \(\psi_{a} = \ket{a}\ket{b}\ket{0}\), with \(\ket{a}\) and \(\ket{b}\) representing the quantum states of the integers to be added. The operator facilitates the modular addition, transitioning to the final state \(\psi_{b}=\ket{a}\ket{b}\ket{a + b \mod{N}}\), which embodies the result \(\ket{a + b \mod{N}}\) stored in the third register.}
    \label{fig:quantum_quantum_outplace_sum}
\end{figure}

The schematic reveals a white-and-black diagonal control point that initiates a sequence of controlled gates (refer to Fig. \eqref{fig:quantum_quantum_outplace_sum}). The control point depicted employs a series of controlled gates, with each gate's action scaled down by powers of two by the binary sequence of the input qubits. This ensures a precise manipulation of the quantum states, which is crucial for the intended computational process.

This operator is conceptually constructed upon the \(Add_{\text{out}}(k, N)\) operator, harnessing the \textit{Inplace} modular adder method. The evolution of states through this operator is represented in equation \eqref{eq:outplace_seq}:
\begin{equation}
    \ket{a}\ket{b}\ket{0} \xrightarrow{\text{Copy}} \ket{a}\ket{b}\ket{a} \xrightarrow{Add_{\text{in}}(N)} \ket{a}\ket{b}\ket{a+b \mod N}.
    \label{eq:outplace_seq}
\end{equation}

For a visual depiction of the circuit effectuating this transformation, refer to Fig. \eqref{fig:quantum_quantum_outplace_sum}.

\subsection{\textit{Outplace} Modular Quantum-Classical Multiplier}
The \textit{Outplace} Modular Quantum-Classical multiplier, denoted as \(Mult_{\text{out}}(k,N)\), is defined by the equation \eqref{eq:mult_q_c_out}.
\begin{equation}
    Mult_{\text{out}}(k,N)\ket{a}\ket{b} = \ket{a}\ket{b+ka \mod N}
    \label{eq:mult_q_c_out}
\end{equation}

To construct this operator, we transition the qubit \(\ket{b}\) into the Fourier basis. Subsequently, controlled summations are applied, represented as \(c_{\ket{a_i}}Add_{\text{in}}(k2^{n-i-1}, N)\ket{b}\). 

\begin{figure*}
    \centering
    \resizebox{1\textwidth}{!}{
    \includegraphics[width=1\textwidth]{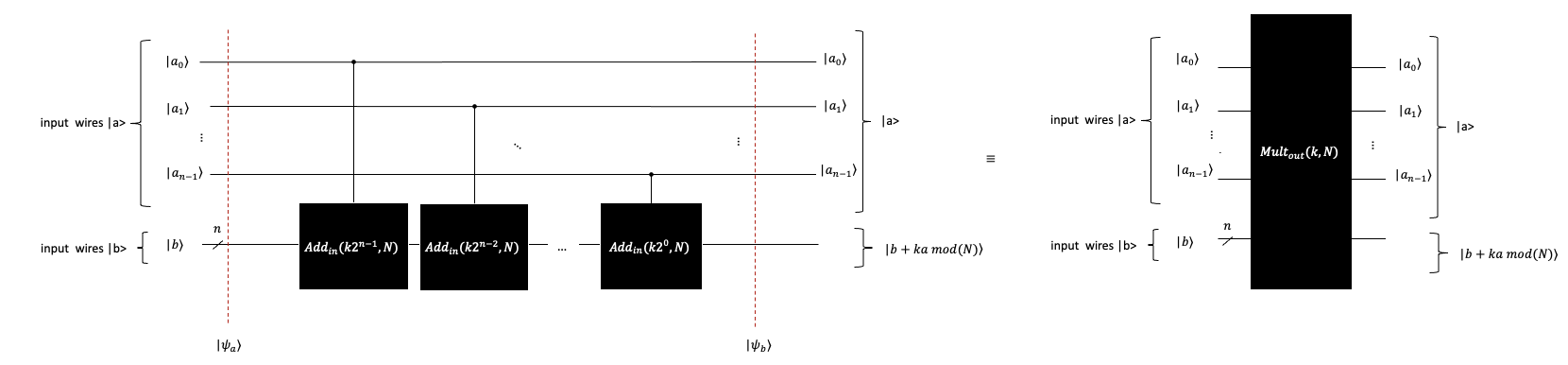}
    }
    \caption{Depiction of the \textit{Outplace} Modular Quantum-Classical Multiplier Operator. The operation commences with the state \(\psi_{a} = \ket{a}\ket{b}\), which, through the multiplier, transitions to \(\psi_{b}=\ket{a}\ket{b + ka \mod{N}}\). This transition embodies the modular multiplication by a classical constant \(k\), pivotal in the computational steps of the \textit{Inplace} Quantum-Classical Multiplier.}

    \label{fig:quantum_classic_outplace_mult_fig}
\end{figure*}

The representation of this transformation within a quantum circuit is showcased in Fig. \eqref{fig:quantum_classic_outplace_mult_fig}.

\subsection{\textit{Inplace} Modular Quantum-Classical Multiplier}\label{sec:inplace_q_c_multiplier}

\begin{figure*}
    \centering
    \resizebox{1\textwidth}{!}{
    \includegraphics[width=1\textwidth]{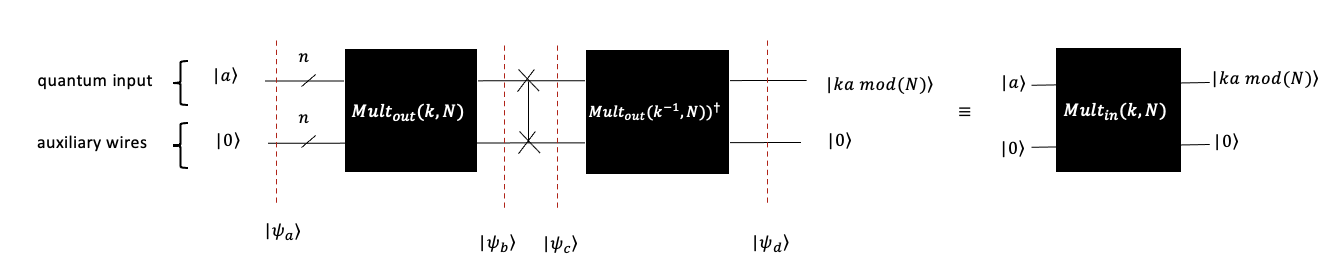}
    }
    \caption{The \textit{Inplace} Modular Quantum-Classical Multiplier Circuit initiates with the state \(\psi_a = |a\rangle|0\rangle\), transitioning to the intermediate state \(\psi_b = |a\rangle|ka\rangle\). Upon verification that \(k\) and \(N\) are coprime, the circuit implements the transformation to \(\psi_c = |ka\rangle|a-k^{-1}ka\rangle\), culminating in the final output state \(\psi_d = |ka \mod N\rangle\), which reflects the desired \textit{Inplace} quantum modular multiplier.}

    \label{fig:quantum_classic_inplace_mult_fig}
\end{figure*}

The \textit{Inplace} Modular Quantum-Classical Multiplier (see Fig. \eqref{fig:quantum_classic_inplace_mult_fig}), denoted as \(Mult_{\text{in}}(k, N)\), is defined by the transformation in equation \eqref{eq:quantum_classic_outplace_mult}:
\begin{equation}
    Mult_{\text{in}}(k, N)\ket{a} = \ket{k a \mod N}.
    \label{eq:quantum_classic_outplace_mult}
\end{equation}

The operator is implementable exclusively when \( k \) possesses a \textbf{modular inverse} relative to \( N \). This constraint is vital since the operator is not universally bijective. For illustration, given \( N = 8 \), one discerns that \( Mult_{\text{in}}(4, 8)\ket{2} = \ket{0 \mod 8} \) and \( Mult_{\text{in}}(4, 8)\ket{4} = \ket{0 \mod 8} \). The primary requisite for this operator's instantiation is the existence of a modular inverse \( k^{-1} \) for \( k \), guaranteeing its bijectiveness.

With the preconditions met and using an auxiliary register, one can deploy the \( Mult_{\text{out}}(k) \) operator twice to achieve the transformation delineated in equation \eqref{eq:quantum_classic_inplace_mult}:

\begin{align}
    \ket{a}\ket{0} & \xrightarrow{Mult_{\text{out}}(k,N)} \ket{a}\ket{ka} \\
    & \xrightarrow{\text{SWAP}} \ket{ka}\ket{a} \\
    & \xrightarrow{{Mult_{\text{out}}(k^{-1}, N)}^{\dagger}} \ket{ka}\ket{0}.
    \label{eq:quantum_classic_inplace_mult}
\end{align}

\subsection{\textit{Outplace} Modular Quantum-Quantum Multiplier}\label{sec:outplace_q_q_multiplication}

\begin{figure*}
    \centering
    \resizebox{1\textwidth}{!}{
    \includegraphics[width=1\textwidth]{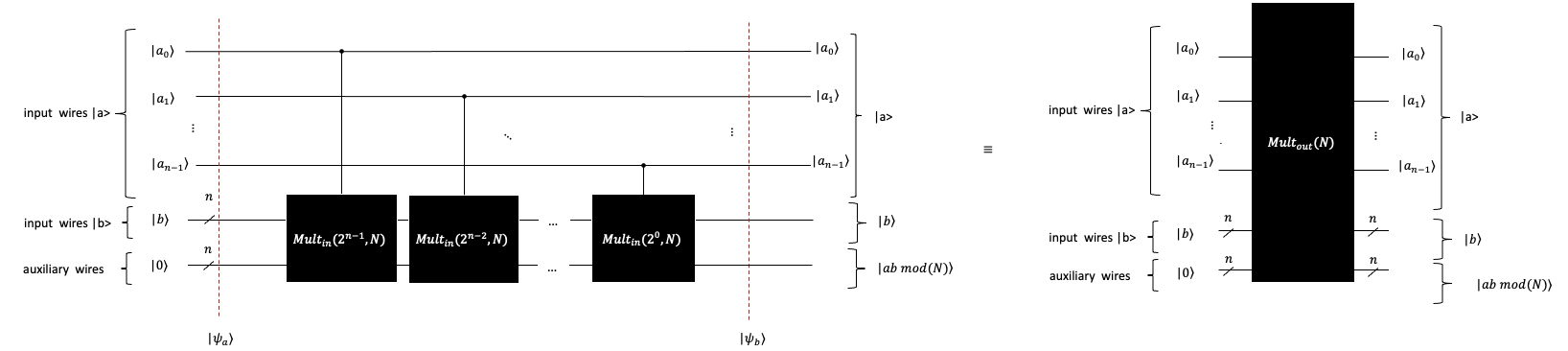}
    }
    \caption{Schematic Illustration of the \textit{Outplace} Modular Quantum-Quantum Multiplier Operator. The process initiates with the state \(\psi_{a} = \ket{a}\ket{b}\ket{0}\), evolving to the outcome \(\psi_{b}=\ket{a}\ket{b}\ket{ab \mod{N}}\), thereby executing the multiplication operation. This operator is a quintessential example of quantum-quantum interaction, efficiently facilitating the multiplication of two quantum states within a modular framework.}
    \label{fig:quantum_quantum_outplace_mult}
\end{figure*}

The \textit{Outplace} Modular Quantum-Quantum Multiplier (see Fig. \eqref{fig:quantum_quantum_outplace_mult}), represented as \(Mult_{\text{out}}(N)\), is characterized by the transformation in equation \eqref{eq:mul_qq_out}.
\begin{equation}
    Mult_{\text{out}}(N)\ket{a}\ket{b}\ket{0} = \ket{a}\ket{b}\ket{ab \mod N}.
    \label{eq:mul_qq_out}
\end{equation}

The underlying mechanism for this operator involves interchanging the third register in the sequential execution of the operations \(c_{a_i}-Mult_{\text{out}}(2^{n-i-1}, N)(\ket{b})\).

\subsection{\textit{Inplace} Modular Quantum-Quantum Multiplier}\label{sec:inplace_qq_mult}

The objective of the \textit{Inplace} Modular Quantum-Quantum Multiplier, denoted by \(Mult_{\text{in}}(N)\), is to execute the transformation described in Eq.~\eqref{eq:mul_qq_in}.
\begin{equation}
    Mult_{\text{in}}(N)\ket{a}\ket{b} \rightarrow \ket{a}\ket{ab \mod N}.
    \label{eq:mul_qq_in}
\end{equation}

To conceive this operator, inspiration can be drawn from the strategy delineated for in-place quantum-classic multiplication, as specified in Eq.~\eqref{eq:quantum_classic_outplace_mult}.

The procedure for the \textit{Inplace} Modular Quantum-Quantum multiplier unravels as a series of state transformations. Commencing with the initial state of \(\ket{a}\ket{b}\ket{0}\ket{0}\), where we encapsulate the product \(ab\) in the third register. This step engenders the intermediate state \(\ket{a}\ket{b}\ket{ab}\ket{a^{-1}}\), as depicted in the progression sequence of Eq.~\eqref{eq:progression_mul_qq_in}.

Subsequent operations, as elucidated in the same sequence, transfer the product from the auxiliary to the principal register, and further transformations culminate in the final state \(\ket{a}\ket{ab}\ket{0}\ket{0}\).

\begin{equation}
    \begin{aligned}
        &\ket{a}\ket{b}\ket{0}\ket{0} \\
        &\quad \xrightarrow{\text{Multiplication and Inverse}} \ket{a}\ket{b}\ket{ab}\ket{a^{-1}} \\
        &\quad \xrightarrow{\text{State Swap}} \ket{a}\ket{ab}\ket{b}\ket{a^{-1}} \\
        &\quad \xrightarrow{\text{Modular Correction}} \ket{a}\ket{ab}\ket{b-a^{-1}ab}\ket{a^{-1}} \\
        &\quad \xrightarrow{\text{Auxiliary Reset}} \ket{a}\ket{ab}\ket{0}\ket{0}
    \end{aligned}
    \label{eq:progression_mul_qq_in}
\end{equation}

While the construction of the \textit{Inplace} Quantum-Quantum Multiplier could theoretically follow a methodology akin to that of the \textit{Inplace} Quantum-Classic Multiplier, such an endeavor has been pragmatically set aside. This decision is attributed to the computational intricacies associated with the modular inverse operation, as encapsulated by \( U \ket{a}\ket{0} = \ket{a}\ket{a^{-1} \mod N} \). The formulation of this inverse operator represents a significant computational challenge that would necessitate a disproportionate allocation of resources and algorithmic complexity, potentially overshadowing the operator's practical utility in the current quantum computing paradigm.

\subsection{\textit{Outplace} Modular Exponential Operator}\label{sec:mod_exp_operator}

\begin{figure*}[!t]
    \centering
    \includegraphics[width=1\textwidth]{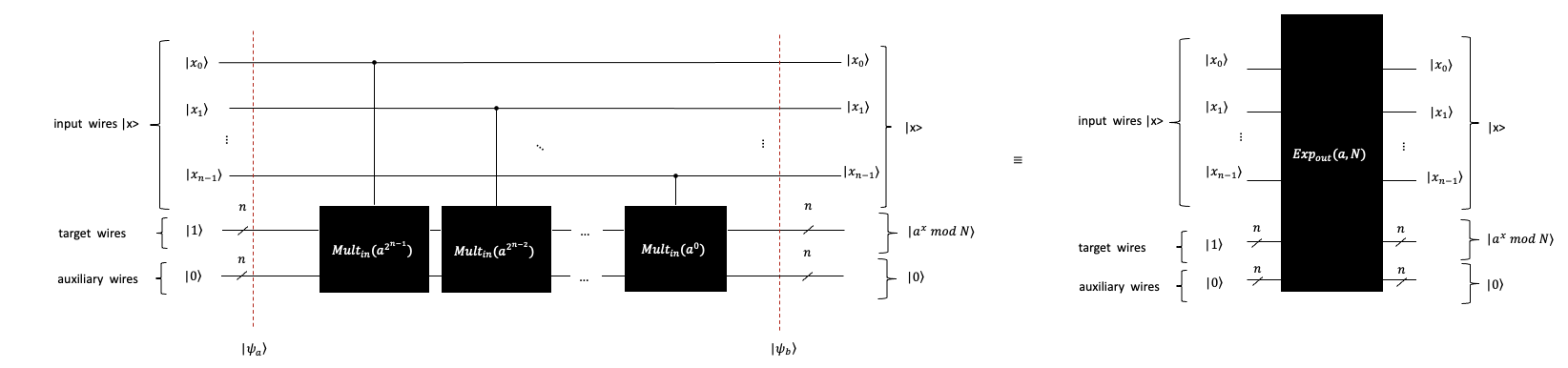}
    \caption{Schematic Depiction of the Quantum Exponential Operator. Beginning as depicted in the figure with the initial state \(\psi_{a} = \ket{x}\ket{1}\ket{0}\), the circuit culminates in the final state \(\psi_{b}=\ket{x}\ket{a^x \mod{N}}\ket{0}\), implementing the operation. The transformation is achieved through an \textit{Inplace} Modular multiplier operator that marries quantum and classical computation in a seamless quantum circuit execution.}

    \label{fig:mod_exp_operator}
\end{figure*}

Central to the machinery of Shor's algorithm is the modular exponential operator (see Fig. \eqref{fig:mod_exp_operator}), which undertakes the transformation delineated in Eq.~\eqref{eq:mod_exp_operator}.
\begin{equation}
    Exp(a,N)\ket{x}\ket{1}\ket{0} \rightarrow \ket{x}\ket{a^x \mod N}\ket{0}.
    \label{eq:mod_exp_operator}
\end{equation}

Considering this operator's pivotal role in quantum algorithmic design, it is imperative to architect its efficient and accurate realization. The circuit's elegance lies in its modular construction, which aligns with the architectural constraints and operational paradigms of current quantum computational frameworks.

The modular exponential operator, integral to executing Shor's algorithm, demands an effective implementation strategy. The foundation of our approach to crafting this operator is expounded through the exploitation of the exponentiation process's binary decomposition, as shown in Eq.~\eqref{eq:exp_decomp}:

\begin{equation}
    a^x = a^{\sum_{i=0}^{n-1} x_i 2^{n-i-1}} = \prod_{i=0}^{n-1} (a^{2^{n-i-1}})^{x_i}. \label{eq:exp_decomp}
\end{equation}

In this algorithmic formulation, the \( n \) necessary values \( a^{2^{n-i-1}} \mod N \) are precomputed, setting the stage for the subsequent construction of the product over indices \( i \) corresponding to \( x_i = 1 \). The ensuing application of the operators \( Mult_{\text{in}}(k, N) \) composes the desired modular exponentiation, as detailed in Eq.~\eqref{eq:exp_decomp}.

To further elucidate the modular exponential operator's workings, we refer to its corresponding quantum circuitry, depicted in Fig.~\ref{fig:mod_exp_operator}. The circuit employs a sequence of controlled modular multiplications, adeptly processing each qubit of the state \(\ket{x}\) to effectuate the prescribed transformation. This systematic approach not only showcases algorithmic precision but also underscores the adaptability and scalability of our design to the constraints of near-term quantum devices.

\section{Results}\label{sec:results}

Our investigation culminates in the precise articulation of quantum modular arithmetic operators, each with detailed implementation strategies and performance metrics. Central to our exposition is the \textit{Outplace} Exponential Operator, whose intricacies are unraveled in Section~\eqref{sec:mod_exp_operator}, and whose operational schema is depicted in Figure~\eqref{fig:mod_exp_operator}. This advanced operator, essential for algorithms like Shor's, is presented with a comprehensive and pragmatic implementation guide, serving as an indispensable blueprint for quantum computing practitioners.

The quintessence of our results is captured in Table \ref{tab:operator_results}, which synthesizes the resource efficiency of the delineated operators. It quantifies their computational demands regarding qubit count, auxiliary qubit, and algorithmic depth. This critical appraisal enables a nuanced understanding of the operators' practical viability and scalability in the quantum computing milieu.

\begin{table}[ht]
\centering
\caption{Resource assessment of quantum operators for modular arithmetic}
\label{tab:operator_results}
\begin{tabular}{lccl}
\toprule
\hline
                Operator & \#qubits & Auxiliary qubits &  Depth \\
\hline
\midrule
  $Add_{\text{in}}(k,N)$ &             $O(n)$ &             $O(1)$ &   $O(n)$ \\
 $Add_{\text{out}}(k,N)$ &             $O(n)$ &             $O(1)$ &   $O(n)$ \\
    $Add_{\text{in}}(N)$ &             $O(n)$ &             $O(1)$ &   $O(n)$ \\
   $Add_{\text{out}}(N)$ &             $O(n)$ &            $O(1)$ &   $O(n)$ \\
$Mult_{\text{out}}(k,N)$ &             $O(n)$ &             $O(1)$ & $O(n^2)$ \\
 $Mult_{\text{in}}(k,N)$ &             $O(n)$ &             $O(1)$ & $O(n^2)$ \\
  $Mult_{\text{out}}(N)$ &             $O(n)$ &             $O(1)$ & $O(n^2)$ \\
              $Exp(a,N)$ &             $O(n)$ &             $O(n)$ & $O(n^3)$ \\
\bottomrule
\hline
\end{tabular}
\end{table}

In tandem with the theoretical insights, we democratize access to our implementations through the open-source release of our code, as delineated in Section~\eqref{sec:code}. The accessibility of our codebase invites replication, validation, and further innovation, fueling the advancement of quantum computational research.

Our work lays the cornerstone for next-generation quantum algorithms. The synergy between rigorous implementation guidance and readily available code catalyzes a broader engagement with quantum computing, lowering the threshold for entry into this specialized domain. Such contributions are pivotal, particularly as we navigate the complexities and possibilities of the quantum computing renaissance.

In adapting the methodologies and quantum circuits detailed in this study thoroughly examined, such as \textit{Qiskit}, we strongly recommend that users carefully consider the encoding and representation of qubits. The present work utilizes a \textit{big-endian} encoding scheme, which must be taken into account to guarantee the correct operation and translatability of our implementations across diverse quantum computing ecosystems.

\section{Conclusion} \label{sec:conclusions}
Our research delineates the pivotal role of quantum operations in advancing computational arithmetic, marking a cornerstone in the \textit{ISQ} era \cite{FromNISQtoISQs}. We have thoroughly examined a spectrum of quantum operators, ranging from adders to the modular exponential operator, which is fundamental to executing advanced quantum algorithms. Our analysis maps the current state of the art and dissects the intricate challenges inherent in scaling quantum computations.

The implications of this work extend beyond mere technical advancements, necessitating a reevaluation of quantum arithmetic's foundational principles. The clear insights presented here offer a vantage point on the trajectory of quantum computation, foreshadowing a new wave of innovations poised to redefine the limits of computational performance. As the quantum paradigm shifts, the strategic insights garnered are indispensable for effectively leveraging quantum mechanics in computational sciences.

In quantum computation, the exponential operator introduced here is paramount, particularly within the subroutine of Shor's algorithm. It is meticulously designed to function with a \(3n\)-qubit system, striking a balance between computational efficiency and the constraints of current quantum processors. This optimization emphasizes practicality, albeit at the expense of direct reversibility, which is a calculated concession given the nascent stage of quantum technologies.

\begin{figure}[!ht]
    \centering
    \resizebox{.5\textwidth}{!}{
    \includegraphics[width=1\textwidth]{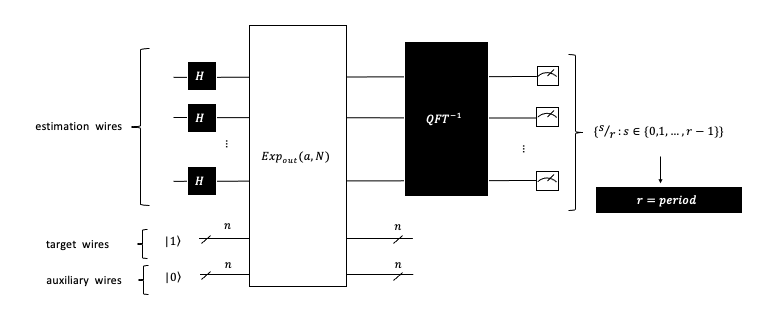}
    }
    \caption{Transformation of the Exponential Operator into the \textit{Period-Finding} \cite{shor1994algorithms} Component for Shor's Algorithm. This diagram delineates the adaptation of the exponential operator to form the crux of the period-finding subroutine. The input qubits are initially prepared in a superposition state by applying Hadamard gates, setting the stage for phase estimation. This is succeeded by the deployment of the inverse Quantum Fourier Transform (\(QFT^{-1}\)) across these qubits. A measurement post-\(QFT^{-1}\) extraction finalizes the period-finding sequence, which is instrumental in the factorization process inherent to Shor's algorithm \cite{shor1994algorithms}.}

    \label{fig:period_finding}
\end{figure}

Our exponential operator, by initializing its inputs in superposition using Hadamard gates and then applying the inverse Quantum Fourier Transform (\(QFT^{-1}\)) to the identical qubits, is effectively converted into the period-finding circuit integral to Shor's algorithm as shown in Fig. \eqref{fig:period_finding}.

The adoption of the \textit{Iterative Quantum Phase Estimation} (IQPE) \cite{dobvsivcek2007arbitrary} technique, as corroborated by extant literature \cite{willsch2023large}, suggests a significant reduction of qubit resources. In particular, the application of \textit{IQPE} is projected to streamline the qubit requirement from \(3n\) down to \(2n+2\), optimizing the quantum circuitry for modular arithmetic operations within a more resource-constrained quantum computational framework.

This convergence of an advanced exponential operator design with the precision of phase estimation algorithms indicates a sophisticated approach toward achieving potential advantage. The focus on the exponential operator does not diminish the collective significance of the suite of quantum operators discussed; instead, it exemplifies the critical innovations necessary for advancing the field of quantum algorithms.

\section*{Code}\label{sec:code}
The Python code implementing the seven modular operators discussed herein is available in the accompanying GitHub repository. The code provides the foundational algorithms enabling in-depth exploration and adaptation of the operators. The code can be accessed at \url{https://github.com/pifparfait/Efficient-Quantum-Modular-Arithmetics}.

\section*{Acknowledgements} The authors thank Guillermo Alonso de Linaje for the discussions and consideration during the experiments.

\section*{Compliance with Ethics Guidelines}

Funding: This research received no external funding. 
Institutional review: This article contains no studies with human or animal subjects.
Informed consent: Informed consent was obtained from all participants in the study.
Data availability: Data sharing is not applicable. No new data were created or analyzed in this study. Data sharing does not apply to this article.

\newpage

\bibliography{main} 
\end{document}